\documentclass{article}

\usepackage[preprint, nonatbib]{nips_2018}

\usepackage{float}

\usepackage[square,numbers]{natbib}
\usepackage{marvosym}
\usepackage[utf8]{inputenc} 
\usepackage[T1]{fontenc}    
\usepackage{hyperref}       
\usepackage{url}            
\usepackage{booktabs}       
\usepackage{amsfonts}       
\usepackage{nicefrac}       
\usepackage{microtype}      
\usepackage{graphicx}
\usepackage{tabularx}
\usepackage{hyperref}
\usepackage{amssymb}
\usepackage{amsmath}
\usepackage{array, colortbl}
\usepackage{diagbox}
\usepackage{longtable}

\definecolor{ruleblue}{rgb}{0.02, 0.52, 0.74}
\definecolor{rowheadblue}{rgb}{0.18, 0.33, 0.5}
\definecolor{rowblue}{rgb}{0.67, 0.81, 0.93}

\newcommand{\Hline}{
	\hline
	
	\hline
	
	\hline
}

\title{GES Model :Combining Pearson Correlation Coefficient Analysis with 
Multilayer Perceptron}

\author{
  Chunyu Sui \textsuperscript{1,\Letter}\\
  \texttt{suichunyu@hotmail.com} \\
\And
  Xinrui Li \textsuperscript{1}\\
  \texttt{lixinrui@outlook.comm} \\
\And
  Yinghang Song \textsuperscript{1} \\
  \texttt{songyinghang12138@outlook.com} \\
\And
  Sirui Huang \textsuperscript{1} \\
  \texttt{huangsirui@gmail.com} \\
\And
  Yunpeng Zan \textsuperscript{3}\\
  \texttt{zanyunpeng@gmail.com} \\
\And
	School of Computer Science and Technology, Shandong University$^{1}$\\
	School of Control science and Engineering,Shandong University$^{1}$\\
	School of Computer Science and Technology, Shandong University$^{3}$
}

\begin{document}
\maketitle

	\begin{abstract}
		With the development of technological progress, mining on asteroids is 
		becoming a 
		reality\cite{butkevivciene2022sharing}\cite{taylor2022phobos}. This 
		paper focuses on how to distribute asteroid 
		mineral resources in a reasonable way to ensure global equity.
		
		To distribute asteroid resources fairly, 7 primary indicators and 20 
		secondary indicators are introduced to build a mathematical model to 
		evaluate global equity and the weights are given by Analytic Hierarchy 
		Process (AHP). Then Global Equity Score(GES) Model based on 12 primary 
		indicators and 40 secondary indicators is built and TOPSIS method is 
		applied to rank all countries. A t-distribution probability density 
		function is applied to simulate the rate of asteroid mining. The 
		Backward Algorithm is applied to quantitatively measure the impact of 
		changing indicators on global equity. Then Pearson correlation 
		coefficient analysis is conducted for each indicator, and t-test is 
		performed lastly. The results demonstrate that asteroid mining promotes 
		global equity that poor countries can be allocated slightly more 
		mineral resources, and a schedule of the implementation of each measure 
		is given.
		
		To gain more insight, sensitivity analysis is conducted and the results 
		demonstrate that scores vary less than $7\%$. It can be concluded that 
		our GES model have great potential as its robustness, accuracy and 
		strengths.
		
	\end{abstract}

	\section{Introduction}
		\subsection{Problem Background and Restatement}
			With the great progress of science and technology, the asteroid 
			mining is gradually becoming a reality\cite{hellgren2016asteroid}. 
			In recent years, more and more countries have agreed to make outer 
			space benefit whole humanly\cite{ferus2022asteroid}, so how to 
			ensure the equitable distribution of benefits after mining 
			asteroids has become an issue to be discussed by all allied powers.
			
			Before the asteroid mining project is officially conducted, there 
			are many unsettled issues that deserve to be discussed and 
			resolved, including the feasibility of mining on asteroids and the 
			equitable distribution of benefits\cite{hessel2021continuous}. 
			Therefore, it is a critical issue to determine a equitable benefit 
			distribution strategy for the project ensuring the project is 
			carried out while promoting world peace and reducing inequality.
			
			Considering the background, we should address 4 questions in the paper:
			
			{\bfseries Task 1:} First, a global equity definition should be developed. Then, find appropriate indicators and build a model to measure global equity. Next, apply the model to a historical or regional analysis to verify its validity.
			
			{\bfseries Task 2:} Describe the possible future state and vision of the asteroid mining industry. Then analyze the impact on global equity by using the global equity measurement model developed in Task 1.
			
			{\bfseries Task 3:} Improve models to analyze and explore how asteroid mining will affect global equity.
			
			{\bfseries Task 4:} Assuming that UN intends to update its Outer 
			Space Treaty to develop asteroid mining and ensure that asteroid 
			mining benefits all of humanity\cite{xie2022framework}, combine 
			your model and results to propose reasonable policies to ensure 
			that asteroid mining will benefits all of humanity.
		\subsection{Our Work}
			First of all ,we give the definition of global equity and introduce 
			5 primary indicators and 26 secondary indicators to measure the 
			established mathematical model. Next, we use Analytic Hierarchy 
			Process to assign weights to every indicator and perform the 
			consistency test on the weight matrix, and finally the test passed. 
			We calculate the global development balance scores and perform 
			calculations for the last 10 years of 
			data\cite{pashkuro2021asteroid}.
			
			Secondly, we combine the data to provide a vision of asteroid 
			mining's future. Next, we improve our model to take into account of 
			the contribution of science and technology to ensure that "those 
			who contribute more get more".. We use the t-distribution 
			probability density function to simulate the profitability of 
			asteroid mining. Finally, we use TOPSIS to rank the contribution of 
			each country\cite{wang2022principal}, and use the contribution of 
			each country to allocate resources.
			
			After that, we calculate the impact of changing indicators on global equity by using the back propagation algorithm. Next, we calculate the correlation between each indicator and the global development balance score using Pearson correlation coefficient analysis and pass the t-test.
			\begin{figure}[H]
				\centering
				\includegraphics[width=\textwidth, height=9cm]{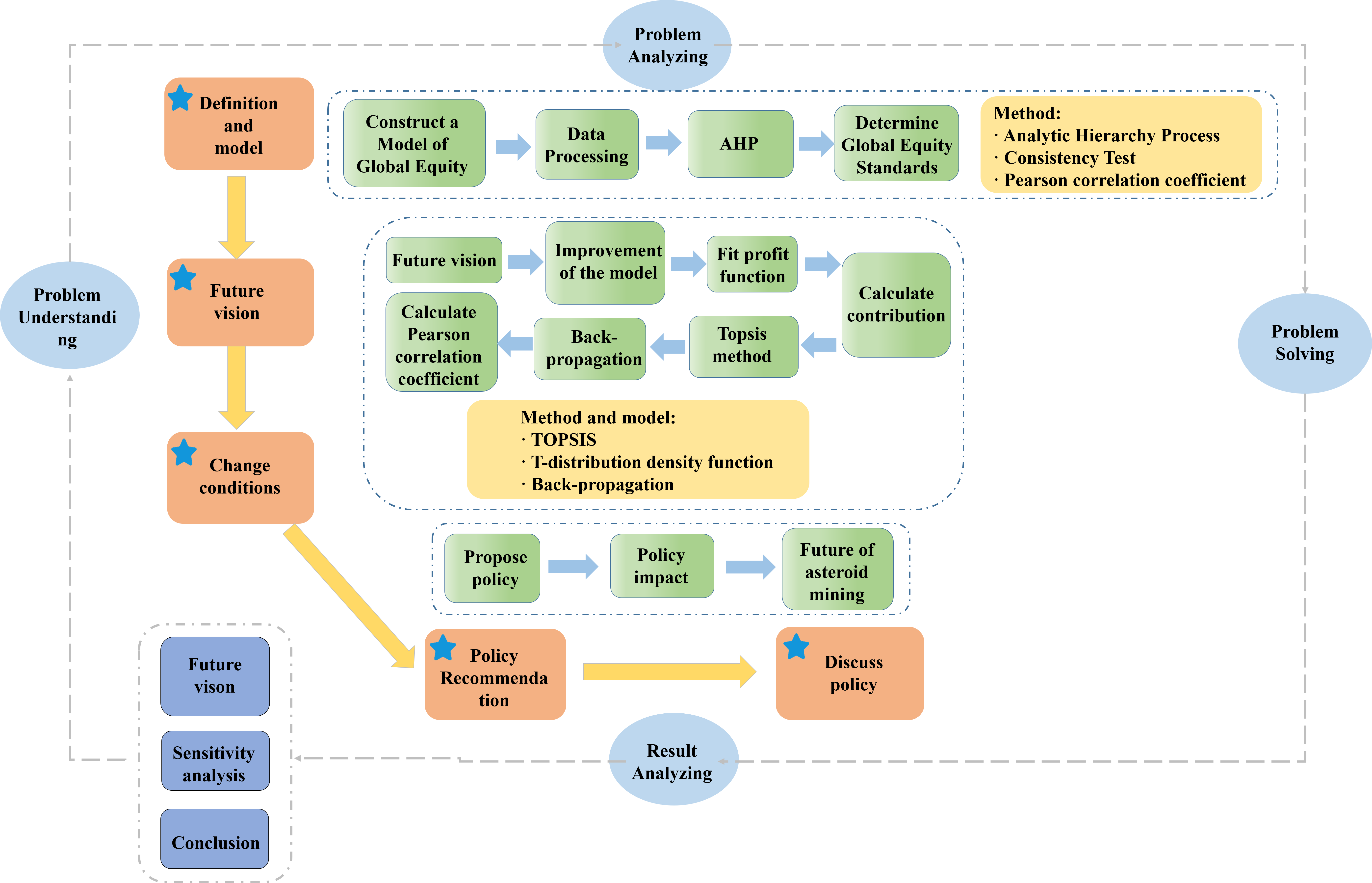}
				\setlength{\abovecaptionskip}{0em}
				\caption{Workflow}
				\label{fig1}
			\end{figure}
	\section{Assumption and Symbol Explanation}
		\subsection{Assumption}
			{\noindent $\bigstar$} Assuming that global equity is influenced by the six indicators including EI, IDG, CEA, MA, HR, ER and SA, and that unexpected factors such as natural disasters do not have obvious impacts on global equity as they may happen anywhere and anytime on the earth.
			
			{\noindent $\bigstar$} Assuming that our mineral extraction rate rises first and then falls, this means we can fit the mineral extraction rate with a t-distribution probability density function curve.
			
			{\noindent $\bigstar$} Assuming stable international conditions and a stable development of the space industry. Meanwhile, mankind can achieve asteroid mining in 15 years or so.
			
			{\noindent $\bigstar$} Assuming that each secondary indicator has a linear effect on the primary indicator, this means that we can easily find the partial derivative of the impact function and analyze the impact of each secondary indicator on global equity accordingly.
		\subsection{Symbol Explanation}
			The primary symbols used in our paper are listed in the table below.
			\begin{table}[H]
			    \centering
			    \label{tab1}
			    \begin{tabular}{p{2cm}<{\centering}p{11cm}<{\centering}}
			         \arrayrulecolor{ruleblue}
    				\Hline
    				\rowcolor{rowheadblue}
    				\textbf{\textit{{\color{white}Variable}}} & \textbf{\textit{{\color{white}Meaning}}} \\
    				\Hline
    				$SA$ & Sustainability \\
    				\Hline
    				\rowcolor{rowblue}
    				$HR$ & Human Resources \\
    				\Hline
    				$CEA$ & Carbon Emission Allocation \\
    				\Hline
    				\rowcolor{rowblue}
    				$EI$ & Economic Indicator \\
    				\Hline
    				$CR$ & Consistency Ratio \\
    				\Hline
    				\rowcolor{rowblue}
    				$CI$ & Consistency Indicator \\ 
    				\Hline
    				$RI$ & Stochastic Consistency Index \\
    				\Hline
    				\rowcolor{rowblue}
    				$Eq_k$ & Development Score of Country k  \\
    				\Hline
    				$GE$ & Global Development Balance Score \\
    				\Hline
    				\rowcolor{rowblue}
    				$w^{(l)}_{jk}$ &
    				\begin{tabular}[c]{@{}c@{}}
    					The Weights Between the $k^{th}$ Neuron in Layer l-1 and the $j^{th}$ Neuron \\ in Layer l
    				\end{tabular} \\
    				\Hline
    				$\boldsymbol{w^{(l)}}$ & Weight Matrix for Layer l-1 to Layer l \\
    				\Hline
    				\rowcolor{rowblue}
    				$b^{(l)}_{j}$ & Bias of the $j^{th}$ Neuron of the $l^{th}$ Layer \\
    				\Hline
    				$\boldsymbol{b^{(l)}}$ & Bias Vector of the $l^{th}$ Layer \\
    				\Hline
    				\rowcolor{rowblue}
    				$z^{(l)}_{j}$ & The Input Value of the $j_{th}$ Neuron of the $l^{th}$ Layer \\
    				\Hline
    				$\boldsymbol{z^{(l)}}$ & The Input Vector of the $l^{th}$ Layer \\
    				\Hline
    				\rowcolor{rowblue}
    				$a^{(l)}_{j}$ & 
    				\begin{tabular}[c]{@{}c@{}}
    					The Activation Value of the $j^{th}$ Neuron of the $l^{th}$ Layer
    				\end{tabular} \\
    				\Hline
    				$\boldsymbol{a^{(l)}}$ & The Activation Output Vector of the $l^{th}$ Layer \\
    				\Hline
    				\rowcolor{rowblue}
    				$N^{(l)}$ & Number of Neurons in Layer l \\
    				\Hline
    				$p(y)$ & T-distribution Probability Density Function \\
    				\Hline
    				\rowcolor{rowblue}
    				$X$ & Forwarding Matrix \\
    				\Hline
    				$Z$ & Normalization Matrix \\
    				\Hline
    				\rowcolor{rowblue}
    				$r$ & Pearson correlation coefficient \\
    				\Hline
			    \end{tabular}
			\end{table}
			
	\section{Global Eauity Score Model}
		\subsection{Defination of Equity}
			To address this issue, first we should define what is equity, "equity means equal rights for all people on earth". More specifically, rights include: fair distribution of resources, fair income, fair opportunities, fair carbon emissions, etc. 
			
			To measure global equity, we adopt the method of calculating the variance after computing the development equity factor for each country. Then, we combine Zhang's descriptions\cite{bib:1} to establish the following model to measure global equality.
		\subsection{Global Equity Evaluation System}
			\subsubsection{Establishment of Evaluation Indicators}
\begin{figure}[H]
					\centering
					\includegraphics[width=11cm, height=10cm]{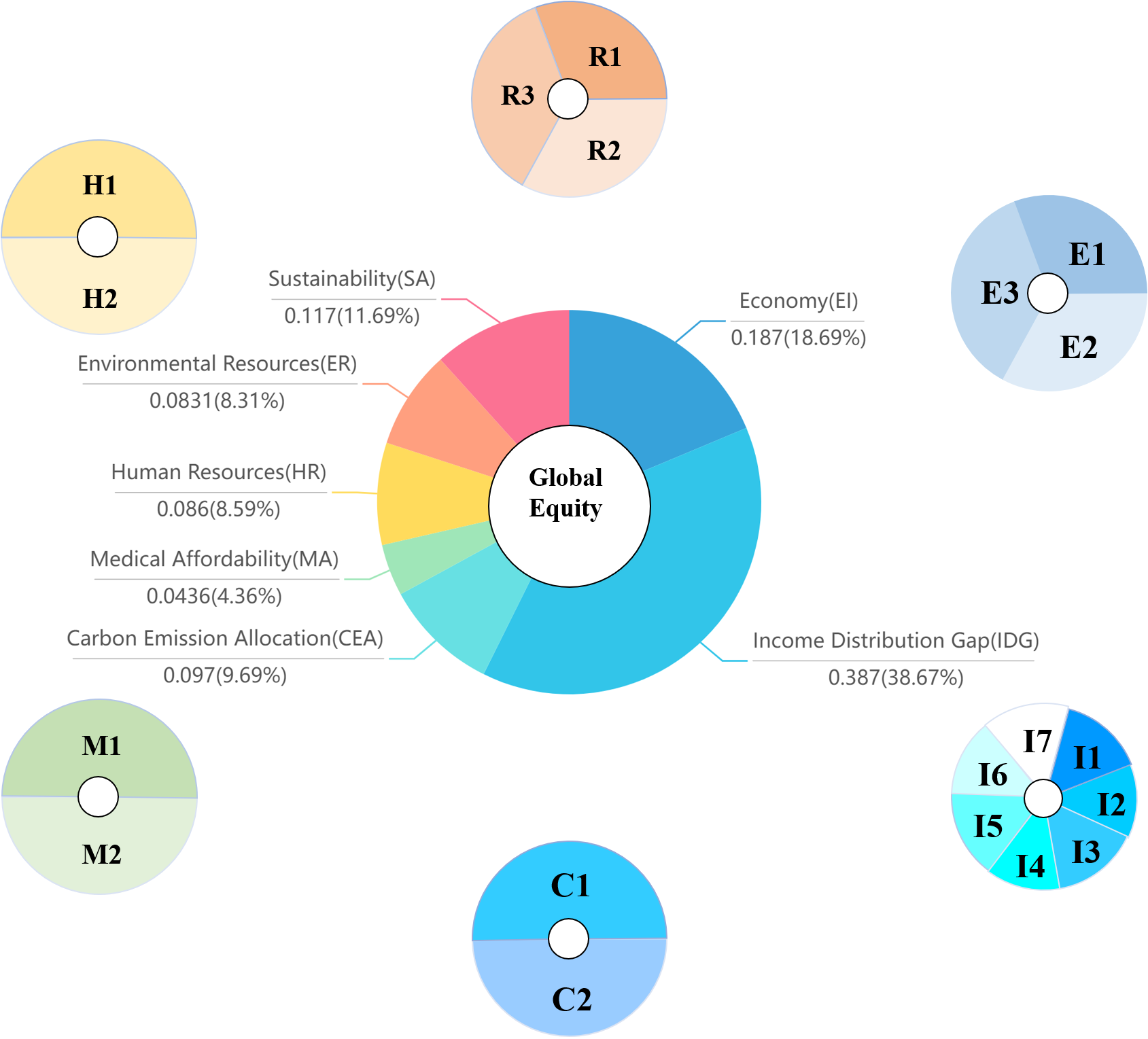}
					\setlength{\abovecaptionskip}{0em}
					\caption{Primary and Secondary Indicators}
					\label{fig2}
				\end{figure}

				We use 7 primary indicators and 15 secondary indicators to measure Country Development Score. Country Development Score for each country is the ratio of the country's development score to the average of the development scores of other countries. The primary indicators that affect Country Development Score are shown in figure \ref{fig2}.
			\subsubsection{AHP Method and Model Validation}
				Next, we use the Analytic Hierarchy Process (AHP) to calculate the weights\cite{shim1989bibliographical}. The weight coefficient matrix of the seven primary indicators is established as follows.
				\begin{figure}[H]
					\centering
					\includegraphics[width=9cm, height=7cm]{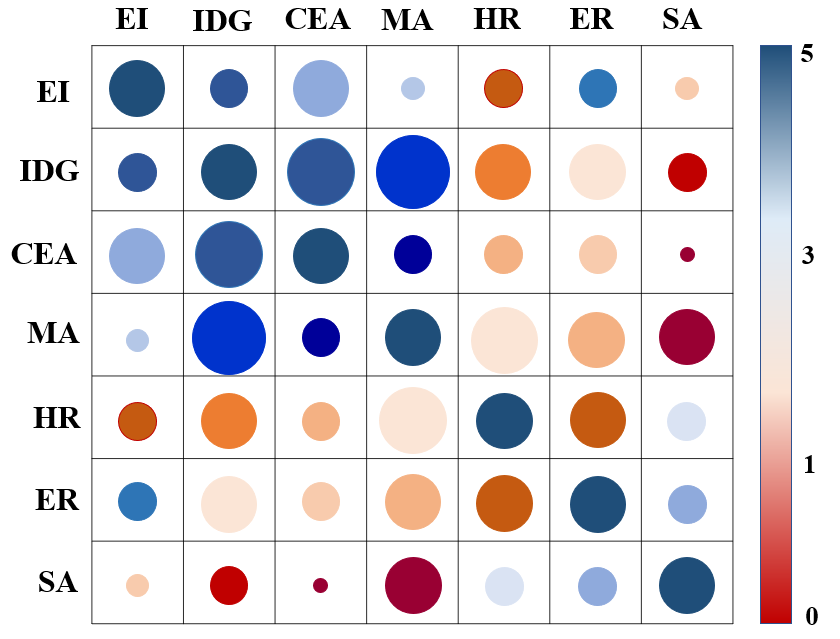}
					\setlength{\abovecaptionskip}{0em}
					\caption{The Weight Coefficient Matrix}
					\label{fig3}
				\end{figure}
				Then we obtain the eigenvalues of the matrix and show it in table \ref{tab2} .
				\begin{table}[H]
					\centering
					\caption{The Eigenvalues of the Weight Coefficient Matrix}
					\label{tab2}
					\arrayrulecolor{ruleblue}
					\begin{tabular}{cp{1.2cm}<{\centering}p{1.2cm}<{\centering}p{1.2cm}<{\centering}p{1.2cm}<{\centering}p{1.2cm}<{\centering}p{1.2cm}<{\centering}p{1.2cm}<{\centering}}
						\Hline
						\rowcolor{rowheadblue}
						{\textbf{\textit{\color{white}Indicators}}}
						& \textbf{\textit{{\color{white}EI}}} & \textbf{\textit{{\color{white}IDG}}} & \textbf{\textit{{\color{white}CEA}}} & \textbf{\textit{{\color{white}MA}}} & \textbf{\textit{{\color{white}HR}}} & \textbf{\textit{{\color{white}ER}}} & \textbf{\textit{{\color{white}SA}}} \\
						\Hline
						Eighenvalues & 7.7200 & 1.8900 & 1.8900 & 0.2948 & 0.2948 & 0.0000 & 0.0000 \\
						\Hline
					\end{tabular}
				\end{table}
				
				{$\divideontimes$ \bfseries{Consistency Check}}
				
				After getting the eigenvalues of the weight coefficient matrix, we perform a consistency check, the equations are as follow
				\begin{equation}
					CI = \frac{\lambda_{max}-n}{n-1} = \frac{7.72 - 7}{7 - 1} = 0.12.
					\label{equ1}
				\end{equation}
			
				Also, RI denotes stochastic consistency index, and its standard values are shown in table \ref{tab3} .
				\begin{table}[H]
					\centering
					\caption{Values of RI}
					\label{tab3}
					\arrayrulecolor{ruleblue}
					\begin{tabular}{p{0.9cm}<{\centering}!{\color{ruleblue}\vrule width 1.5pt}p{0.8cm}<{\centering}p{0.8cm}<{\centering}p{0.8cm}<{\centering}p{0.8cm}<{\centering}p{0.8cm}<{\centering}p{0.8cm}<{\centering}p{0.8cm}<{\centering}p{0.8cm}<{\centering}p{0.8cm}<{\centering}p{0.8cm}<{\centering}}
						\Hline
						\rowcolor{rowheadblue}
						{\textbf{\textit{\color{white}n}}} & {\textbf{\textit{\color{white}1}}} & {\textbf{\textit{\color{white}2}}} & {\textbf{\textit{\color{white}3}}} & 
						{\textbf{\textit{\color{white}4}}} & 
						{\textbf{\textit{\color{white}5}}} &
						{\textbf{\textit{\color{white}6}}} & 
						{\textbf{\textit{\color{white}7}}} & 
						{\textbf{\textit{\color{white}8}}} & 
						{\textbf{\textit{\color{white}9}}} & 
						{\textbf{\textit{\color{white}10}}} \\
						\Hline
						RI & 0.00 & 0.00 & 0.58 & 0.90 & 1.12 & 1.24 & 1.32 & 1.41 & 1.45 & 1.49 \\
						\Hline
					\end{tabular}
				\end{table}
				Substituting the data obtained before, we get
				\begin{equation}
					CR = \frac{CI}{RI} = \frac{0.12}{1.32} = 0.0909 < 0.1.
					\label{equ2}
				\end{equation}
			
\begin{figure}[H]
					\centering
					\setlength{\abovecaptionskip}{-2.5em}
					\includegraphics[width=9cm, height=7cm]{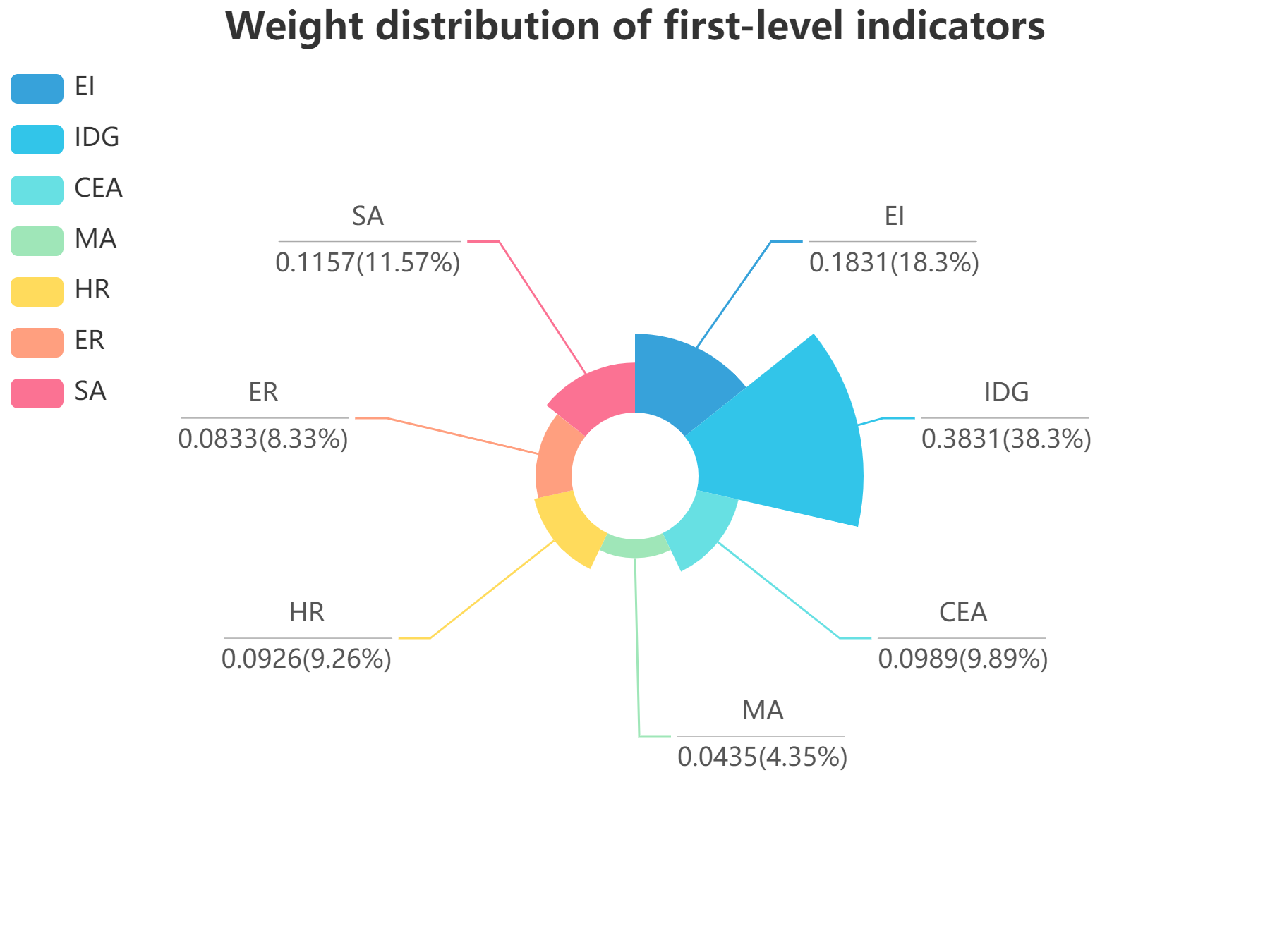}
					\caption{Average Weights}
					\label{fig4}
				\end{figure}

				So it pass the consistency check. Then, we use Arithmetic Mean Method, Geometric Mean Method and Eigenvalue Method to calculate the weight of the primary indicators, finally we get the results shown in table \ref{tab4} .
				\begin{table}[H]
					\centering
					\caption{Weight of the First-level Indicators}
					\label{tab4}
					\arrayrulecolor{ruleblue}
					\begin{tabular}{p{2.2cm}<{\centering} !{\color{ruleblue}\vrule width 1.5pt} p{3.2cm}<{\centering}p{3.2cm}<{\centering}p{3.2cm}<{\centering}}
						\Hline
						\rowcolor{rowheadblue}
						\textbf{\textit{\color{white}
								\begin{tabular}[c]{@{}c@{}}
									Indicators
								\end{tabular}
						}}
						& \textbf{\textit{\color{white}
								\begin{tabular}[c]{@{}c@{}}
									Arithmetic Mean \\ Method
								\end{tabular}
						}}
						& \textbf{\textit{\color{white}
								\begin{tabular}[c]{@{}c@{}}
									Geometric Mean \\ Method
								\end{tabular}
						}} 
						& \textbf{\textit{\color{white}
								\begin{tabular}[c]{@{}c@{}}
									Eigenvalue Method
								\end{tabular}
						}} \\
						\Hline
						EI & 0.1831 & 0.1965 & 0.1810 \\
						\rowcolor{rowblue}
						IDG & 0.3831 & 0.3965 & 0.3810 \\
						CEA & 0.0989 & 0.0996 & 0.0921 \\
						\rowcolor{rowblue}
						MA & 0.0435 & 0.0436 & 0.0438 \\
						HR & 0.0926 & 0.0620 & 0.1027 \\
						\rowcolor{rowblue} 
						ER     & 0.0833 & 0.0852 & 0.0808 \\
						SA     & 0.1157 & 0.1166 & 0.1187 \\
						\Hline
					\end{tabular}
				\end{table}
				Next, after calculating the average of the statistics in table \ref{tab4} , we get figure \ref{fig4} .
				Finally, we get the formula to measure the national development score($Eq_k$):
				\begin{equation}
					\begin{aligned}
						Eq_k &= 0.187 \times EI + 0.387 \times IDG + 0.097 \times CEA \\ &+ 0.0436 \times MA 
						+ 0.086 \times HR + 0.0831 \times ER \\ &+ 0.117 \times SA.
					\end{aligned}
					\label{equ3}
				\end{equation}
			
				{$\divideontimes$ \bfseries{Model Validation}}
				
\begin{figure}[H]
					\begin{minipage}{0.5\linewidth}
						\centerline{\includegraphics[width=7cm, height=4.2cm]{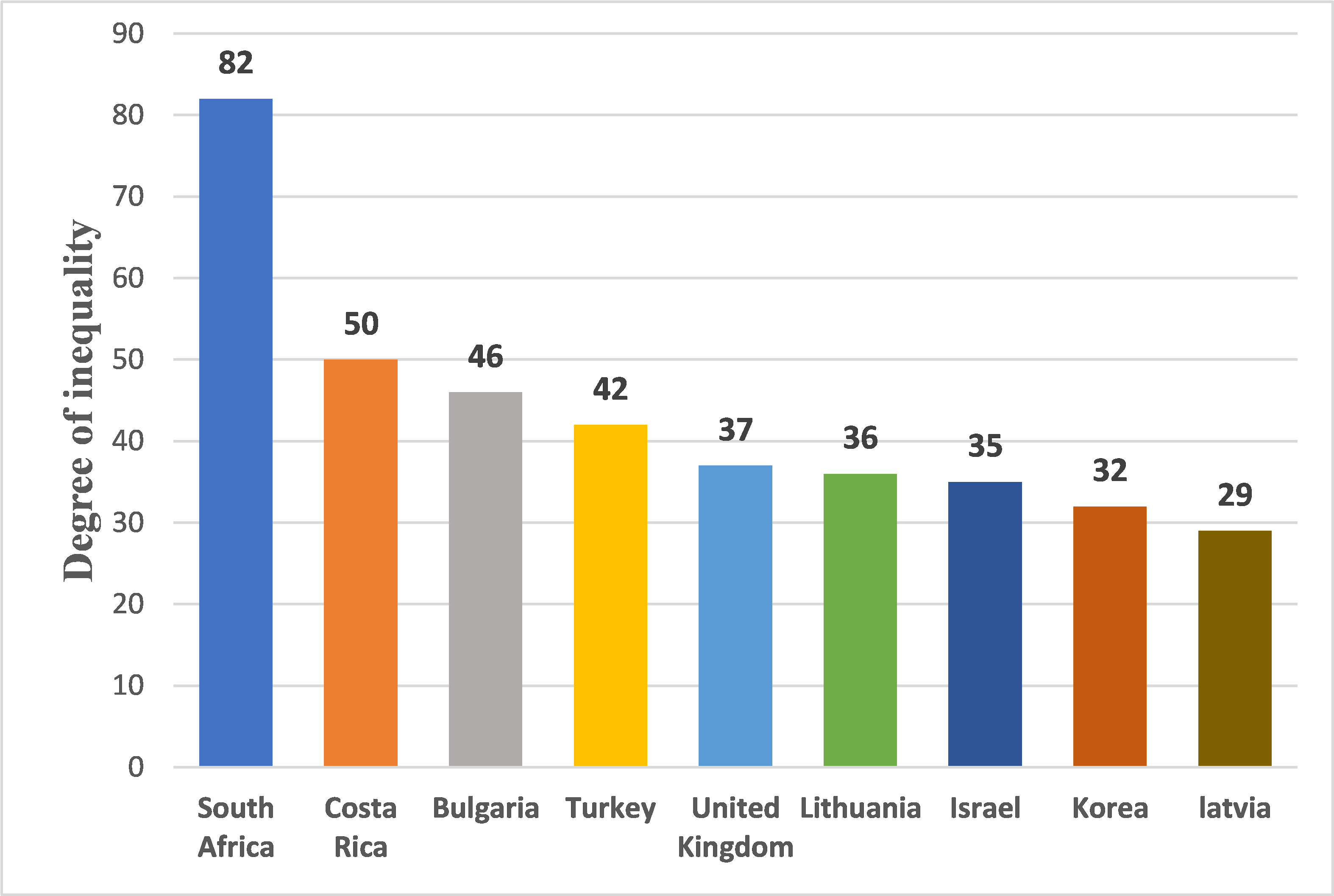}}
						\caption{Top 9 Countries with the Highest Degree of Inequity}
						\label{fig5}
					\end{minipage}
					\hspace{0.2mm}
					\begin{minipage}{0.5\linewidth}
						\centerline{\includegraphics[width=7cm, height=4.2cm]{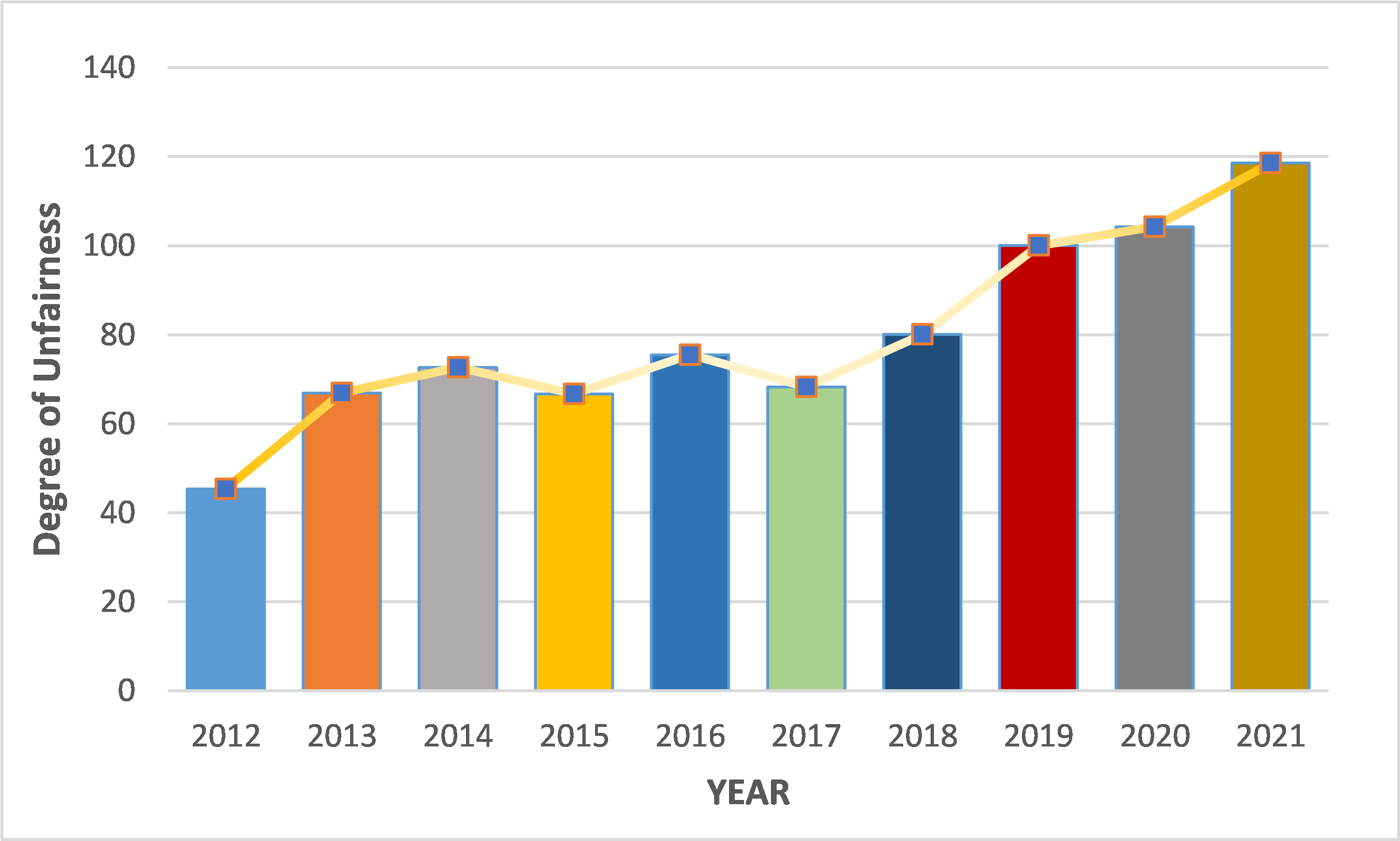}}
						\caption{the Global Inequity Index for 2012 - 2021}
						\label{fig6}
					\end{minipage}
				\end{figure}

				Take Income Distribution Gap(IDG) as an example, according to the factors affecting IDG obtained from the previous analysis, the income inequality score of each country is finally shown in figure \ref{fig5} .
				
				For a certain country, first we de-self the country by comparing its development score($Eq_k$) with the average of other countries' development scores($Eq_m$)($k \neq m$). Then we subtract it from the mean of all selected countries($\overline{Eq_k}$) and take the mean value. That is, using the following formula
				\begin{equation}
					GE = \frac{1}{10n}\sum{(\frac{Eq_k}{\sum\limits_{i \neq k} Eq_i / (n-1)}-\overline{Eq_k})^2}.
					\label{equ4}
				\end{equation}
				
				On balance, we derive the degree of inequity over the last 10 years which was shown in  figure \ref{fig6} .
		\subsection{Our Conclusion}
			By calculating the Global Development Imbalance Index (GDI), it can be concluded that global imbalances are tending to accelerate over the past decade due to factors such as uneven economic development.
	\section{Impacts of Asteroid Mining on Global Equity}
		To explore the impact asteroid mining can have, we first analyze the value asteroids can bring. Take minerals from asteroids as an example, relevant evidence suggests that the asteroid is indeed rich in minerals.
		
		With the depletion of Earth's resources, people are destined to place goals in outer space, and the first bear to brunt is asteroids. In this task, we have made an outlook on the future of asteroid mining and analyze it.
		\subsection{Promoting Global Economic Development}
			Through our research, we found that some of the asteroids have the following commercial valuations shown in table \ref{tab5} .
			\begin{table}[H]
				\centering
				\caption{Commercial Valuations}
				\label{tab5}
				\begin{tabular}{p{2.2cm}<{\centering}p{3.1cm}<{\centering}p{3.1cm}<{\centering}p{3.1cm}<{\centering}}
					\arrayrulecolor{ruleblue}
					\Hline
					\rowcolor{rowheadblue}
					\textbf{\textit{\color{white}Asteroid}} & \textbf{\textit{
							\color{white}
							\begin{tabular}[c]{@{}c@{}}
								Est. Value \\ (US\$billion)
							\end{tabular}
					}} & 
					\textbf{\textit{
							\color{white}
							\begin{tabular}[c]{@{}c@{}}
								Est. Profit \\ (US\$billion)
							\end{tabular}
					}} & 
					\textbf{\textit{
							\color{white}
							$\Delta V(km/s)$
					}} \\
					\Hline
					Didymos & 62 & 16 & 5.162 \\
					\Hline
					\rowcolor{rowblue}
					Anteros & 5570 & 1250 & 5.44 \\
					\Hline
					2001 CC21 & 147 & 30 & 5.636 \\
					\Hline
					\rowcolor{rowblue}
					1992 TC & 84 & 17 & 5.648 \\
					\Hline
				\end{tabular}
			\end{table}
			The growth of global economic has slowed down due to factors such 
			as the COVID-19 outbreak and the declining birth rate of the 
			population\cite{hornsey2022protecting}\cite{shreya2022asteroid}. It 
			is speculated that a relatively small metallic asteroid with a 
			diameter of 1.6 km (1 mile) contains more than \$20 trillion worth 
			of industrial and precious metals\cite{matter2013evidence}. Other 
			studies have shown that rockets can achieve "zero loss" of energy 
			in transit by using solar energy. Other costs are negligible 
			relative to the net profit. The complete utilization of just one 
			1.6 km asteroid could generate over \$20 trillion in net profits. 
			We speculate that after a breakthrough in space technology, the 
			global economy will grow at a rate of at least 9\% per year, and 
			the economy will grow by more than \$10.17 trillion per 
			year\cite{diebold2009measuring}.
		\subsection{Space Becomes a Major Battleground for Human Development}
			The economic appeal of asteroid mining is clear: precious metals 
			such as gold and platinum sell for around US \$50,000 per 
			kilogram\cite{xie2021target}. Due to the huge potential value of 
			planetary mining\cite{dong2020exploration}, space is becoming a 
			major battleground for various countries\cite{ren2022manage}. A 
			breakdown of the first successful missions by country is as 
			follows. At the same time, some strong aviation companies (e.g., 
			SPACEX) will also participate\cite{liu2022tiny}, and the booming 
			aviation business will drive the progress of technology and the 
			growth of the job market\cite{plancher2022tinymledu}.
			\begin{figure}[H]
				\centering
				\includegraphics[width=12cm, height=4.5cm]{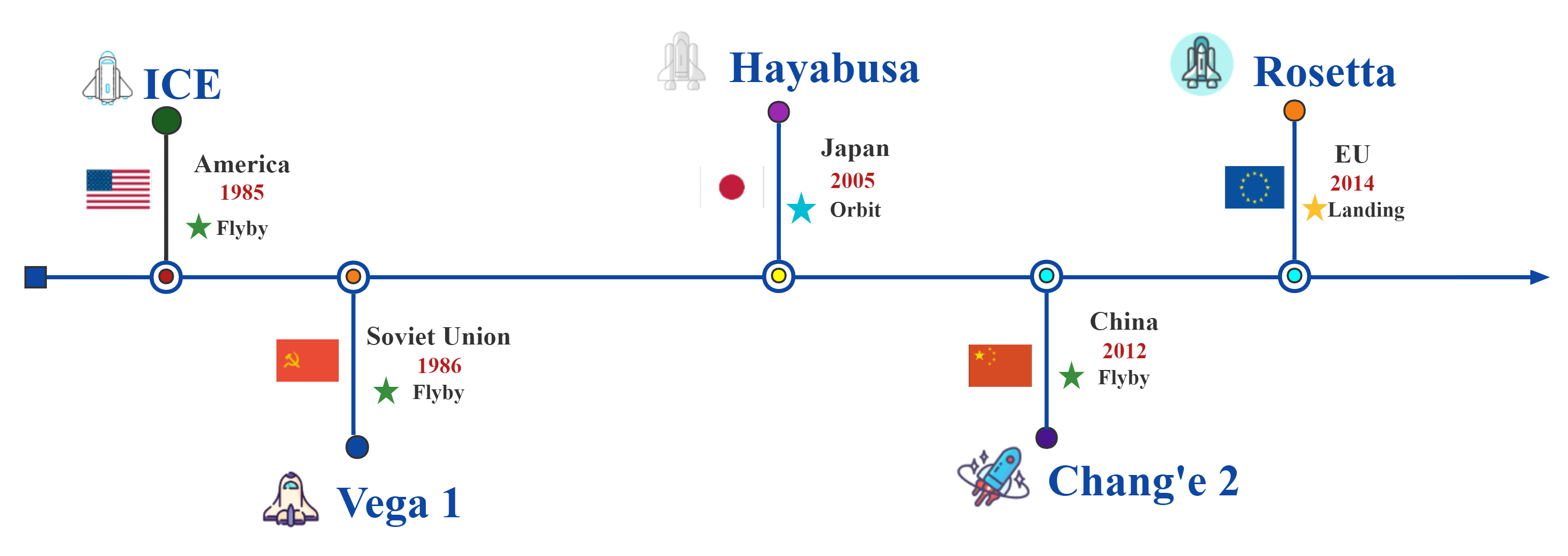}
				\setlength{\abovecaptionskip}{0em}
				\caption{First Successful Mission by Country}
			\end{figure}
		\subsection{Improving the Environment}
			Asteroid mining has equally profound effects on the Earth's 
			environment\cite{gopinath2019compiling}. Take the mining of the 
			rare earth resource platinum, for example. Space mining would have 
			a lower environmental impact\cite{steffen2022explore}, if the 
			spacecraft is able to return 
			between 0.3\% and 7\% of its mass in platinum to Earth, assuming 
			100\% primary platinum or 100\% secondary platinum, 
			respectively.\cite{hein2018exploring}
			
			At the same time, the benefits of asteroid mining are not limited 
			to the replenishment of resources\cite{ye2022asteroid}; it also 
			offers a solution to the greenhouse effect of recent 
			years\cite{dahl2021prospects}.
			Here is some data, we compare the amount of CO2 emissions from asteroid mining to Earth-based mining. Let $b$ denotes kg of payload mass launched into space vs. kg of resources delivered to the target destination, then, from table \ref{tab6} .
			\begin{table}[H]
				\centering
				\caption{Comparison of Space and Earth-based platinum mining greenhouse gas emissions}
				\label{tab6}
				\begin{tabular}{p{2.5cm}<{\centering}cccc}
					\arrayrulecolor{ruleblue}
					\Hline
					\rowcolor{rowheadblue}
					\textbf{\textit{\color{white}$b_{mining}$}} & 
					\textbf{\textit{\color{white}
					\begin{tabular}[c]{@{}c@{}}
						$CO_2 eq$\\ /kg Pt
					\end{tabular}}} & 
					\textbf{\textit{\color{white}
							\begin{tabular}[c]{@{}c@{}}
								Ratio Reference \\
								(40 t / kg $CO_2 eq$) \\
								Earth vs. Space \\
					\end{tabular}}} &
					\textbf{\textit{\color{white}
					\begin{tabular}[c]{@{}c@{}}
						Ratio Reference \\
						(2 t / kg $CO_2 eq$) \\
						Earth vs. Space \\
					\end{tabular}}} &
					\textbf{\textit{\color{white}
					\begin{tabular}[c]{@{}c@{}}
						Earth: 33\% secondary, \\ 66\% primary platinum \\
						vs. Space \\
					\end{tabular}}} \\
					\Hline
					10 & 69 & 580 & 29 & 396 \\
					\Hline
					\rowcolor{rowblue}
					20 & 65 & 620 & 31 & 424 \\
					\Hline
					30 & 63 & 635 & 32 & 434 \\
					\Hline
					\rowcolor{rowblue}
					40 & 62 & 643 & 32 & 439 \\
					\Hline
				\end{tabular}
			\end{table}
			We can see that though asteroid mining account for a lower bound in $CO_2 eq$, compared to Earth-based mining, one order of magnitude higher emissions would lead to one order of magnitude savings\cite{macwhorter2015sustainable}.
		\subsection{The Impact of Asteroid Mining}
			In order to ensure the fairness of resource allocation among 
			different countries in the asteroid mining project, we supplement 
			our model by adding a measure of the scientific and technological 
			contribution of each country\cite{oughan2022public}. At the same 
			time, in order to narrow the global wealth gap and ensure global 
			equity, we allocate resources slightly more to the extremely poor 
			countries.
			\subsubsection{Determination of the Total Contribution}
				Considering the different contributions of different countries in asteroid mining, the countries that contribute more to science and technology deserve to share more resources.
			\subsubsection{Determination of Scientific and Technological Contribution}
				We measure the importance of a country in asteroid mining by 
				its scientific and technological 
				contribution\cite{frost2022project}. We introduce 5 primary 
				indicators and 21 secondary indicators. We use AHP method to 
				calculate the weight of each indicator and the contribution 
				degree is calculated as follows.
				\begin{figure}[H]
					\centering
					\includegraphics[width=\textwidth, height=8cm]{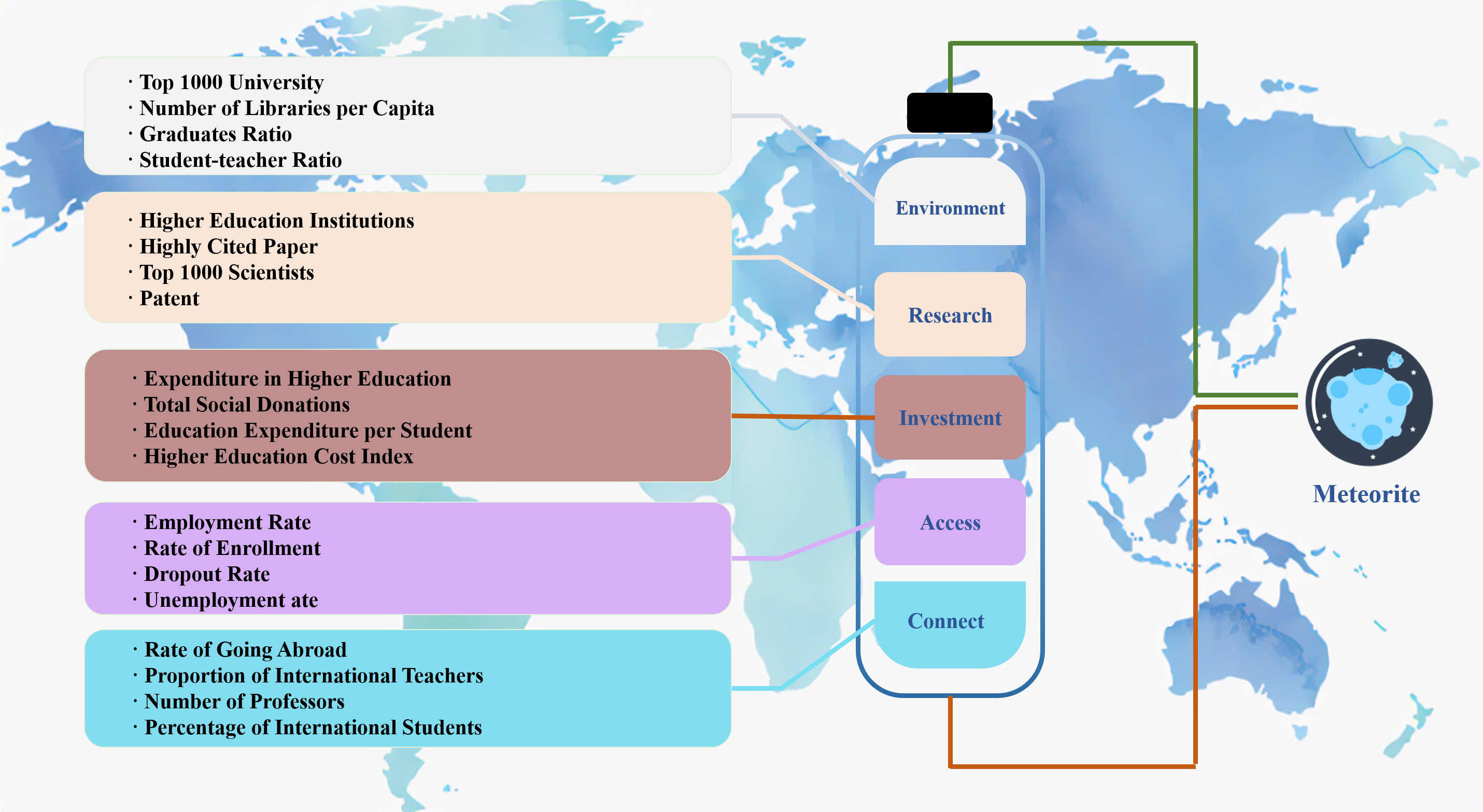}
					\setlength{\abovecaptionskip}{0em}
					\caption{Indicator and the Contribution Degree}
				\end{figure}
			\subsubsection{Calculation Total Score by Using TOPSIS}
				After defining how these variables are determined, we use the TOPSIS method to continue our analysis of the problem\cite{olson2004comparison}.
				
				{$\divideontimes$ \bfseries The basic Steps of TOPSIS}
				
				{$\bullet$ \bfseries Normalize the Original Matrix}
				
				Suppose ${x_i}$ is a set of intermediate type indicator series and the optimal value is $x_{best}$, then equation \ref{equ5} , \ref{equ6} are the forwarding equations.
				\begin{equation}
					M = \max{|x_i - x_{best}|}
					\label{equ5}
				\end{equation}
				\begin{equation}
					\tilde{x_i} = 1 - \frac{|x_i - x_{best}|}{M}
					\label{equ6}
				\end{equation}
				
				{$\bullet$ \bfseries Normalize the Normalization Matrix}
				
				If the normalized matrix is noted as Z, then we use equation \ref{equ7} to normalize the matrix X.
				\begin{equation}
					z_{ij} = x_{ij} / \sqrt{\sum\limits_{i = 1}^{n}x_{ij}^2}.
					\label{equ7}
				\end{equation}
			
				Assume that there are n objects to be evaluated and a standardized matrix of m evaluation indicators.
				\begin{equation}
					\begin{aligned}
						\boldsymbol{Z} &= 
						\begin{bmatrix}
							z_{11} & z_{12} & \cdots & z_{1m} \\
							z_{21} & z_{22} & \cdots & z_{2m} \\
							\vdots & \vdots & \ddots & \vdots \\
							z_{n1} & z_{n2} & \cdots & z_{nm}
						\end{bmatrix} \\
						&=
						\begin{bmatrix}
							0.4056 & 0.2154 & \cdots & 0.2433 \\
							0.3187 & 0.1031 & \cdots & 0.2305 \\
							\vdots & \vdots & \ddots & \vdots \\
							0.0290 & 0.1511 & \cdots & 0.2122 \\
						\end{bmatrix}
					\end{aligned}
					\label{equ8}
				\end{equation}.
				Then we define maximum value($Z^{+}$):
				\begin{equation}
					\begin{aligned}
						Z^{+} &= (Z_{1}^{+}, Z_{2}^{+}, \cdots Z_{m}^{+}) \\
						&= (\max{\{z_{11}, z_{21}, \cdots, z_{n1}\}}, \max{\{z_{12}, z_{22}, \cdots, z_{n2}\}}, \\ &\cdots, \max{\{z_{1m}, z_{2m}, \cdots, z_{nm}\}}),
					\end{aligned}
					\label{equ9}
				\end{equation}
				minimun value($Z^{-}$):
				\begin{equation}
					\begin{aligned}
						Z^{-} &= (Z_{1}^{-}, Z_{2}^{-}, \cdots Z_{m}^{-}) \\
						&= (\min{\{z_{11}, z_{21}, \cdots, z_{n1}\}}, \min{\{z_{12}, z_{22}, \cdots, z_{n2}\}}, \\ &\cdots, \min{\{z_{1m}, z_{2m}, \cdots, z_{nm}\}}),
					\end{aligned}
					\label{equ10}
				\end{equation}
				Distance of the $i^{th}(i = 1, 2, \cdots, n)$ evaluation object from the maximum value($D^{+}_{i}$):
				\begin{equation}
					D^{+}_{i} = \sqrt{\sum\limits_{j = 1}^{m}(Z^{+}_{j} - z_{ij})^2},
					\label{equ11}
				\end{equation}
				Distance of the $i^{th}(i = 1, 2, \cdots, n)$ evaluation object from the minimum value($D^{-}_{i}$):
				\begin{equation}
					D^{-}_{i} = \sqrt{\sum\limits_{j = 1}^{m}(Z^{-}_{j} - z_{ij})^2}.
					\label{equ12}
				\end{equation}
				Then, we can calculate the un-normalized score of the $i^{th}$ evaluation object:
				\begin{equation}
					S_i = \frac{D^{-}_i}{D^{+}_i + D^{-}_i}.
					\label{equ13}
				\end{equation}
				
				Next, we normalize the score using equation \ref{equ14}:
				\begin{equation}
					\tilde{S_i} = S_i / \sum\limits_{i = 1}^{n}S_i.
					\label{equ14}
				\end{equation}
			
				Finally, we calculated the scores of the top 7 countries which was shown in table \ref{tab7} .
				\begin{table}[H]
					\centering
					\caption{Top 7 Overall Scoring Countries}
					\label{tab7}
					\begin{tabular}{cp{1.2cm}<{\centering}p{1.2cm}<{\centering}p{1.2cm}<{\centering}p{1.2cm}<{\centering}p{1.2cm}<{\centering}p{1.2cm}<{\centering}p{1.2cm}<{\centering}}
						\arrayrulecolor{ruleblue}
						\Hline
						\rowcolor{rowheadblue}
						\textbf{\textit{\color{white}Countries}} 
						&
						\textbf{\textit{\color{white}USA}}
						&
						\textbf{\textit{\color{white}China}}
						&
						\textbf{\textit{\color{white}Japan}}
						&
						\textbf{\textit{\color{white}UK}}
						&
						\textbf{\textit{\color{white}France}}
						&
						\textbf{\textit{\color{white}Germany}}
						&
						\textbf{\textit{\color{white}Canada}}
						\\
						\Hline
						Score & 298.9914 & 132.6379 & 64.8017 & 60.4439 & 41.5603 & 34.6638 & 32.9397  \\
						\Hline
					\end{tabular}
				\end{table}
			\subsubsection{Determination of Annual Profit}
				Considering that the technology is not fully mature in the early stage, the mining rate should rise first and then fall. We use the t-distribution probability density function as our mining curve\cite{psarakis1990folded}. The function is defined as follows.
				\begin{equation}
					p(t) = \frac{\Gamma (\frac{n +  1}{2})}{\sqrt{n \pi} \Gamma (\frac{n}{2})}(1 + \frac{y^2}{n})^{-\frac{n+1}{2}}, 0 < t < +\infty
					\label{equ15}
				\end{equation}
				With this definition of the mining curve, we then calculate the income.
				\begin{equation}
					income = \int_{t_1}^{t_2} p'(t) \cdot  V dt
					\label{equ16}
				\end{equation}
				where V denotes total mineral value and we take V = 70 trillion. Then, 
				\begin{equation}
					Profit = income - cost
					\label{equ17}
				\end{equation}
			\subsubsection{Determination of Poor Countries}
				In order to narrow the global wealth gap and ensure global equity, we provide assistance to countries with extreme poverty by giving them slightly more resources. We use GDP as a measure of extreme poverty, and countries in the bottom 20 of the world GDP ranking are considered extremely poor. World GDP per Captica is shown below.
				\begin{figure}[H]
					\centering
					\includegraphics[width=\textwidth, height=8cm]{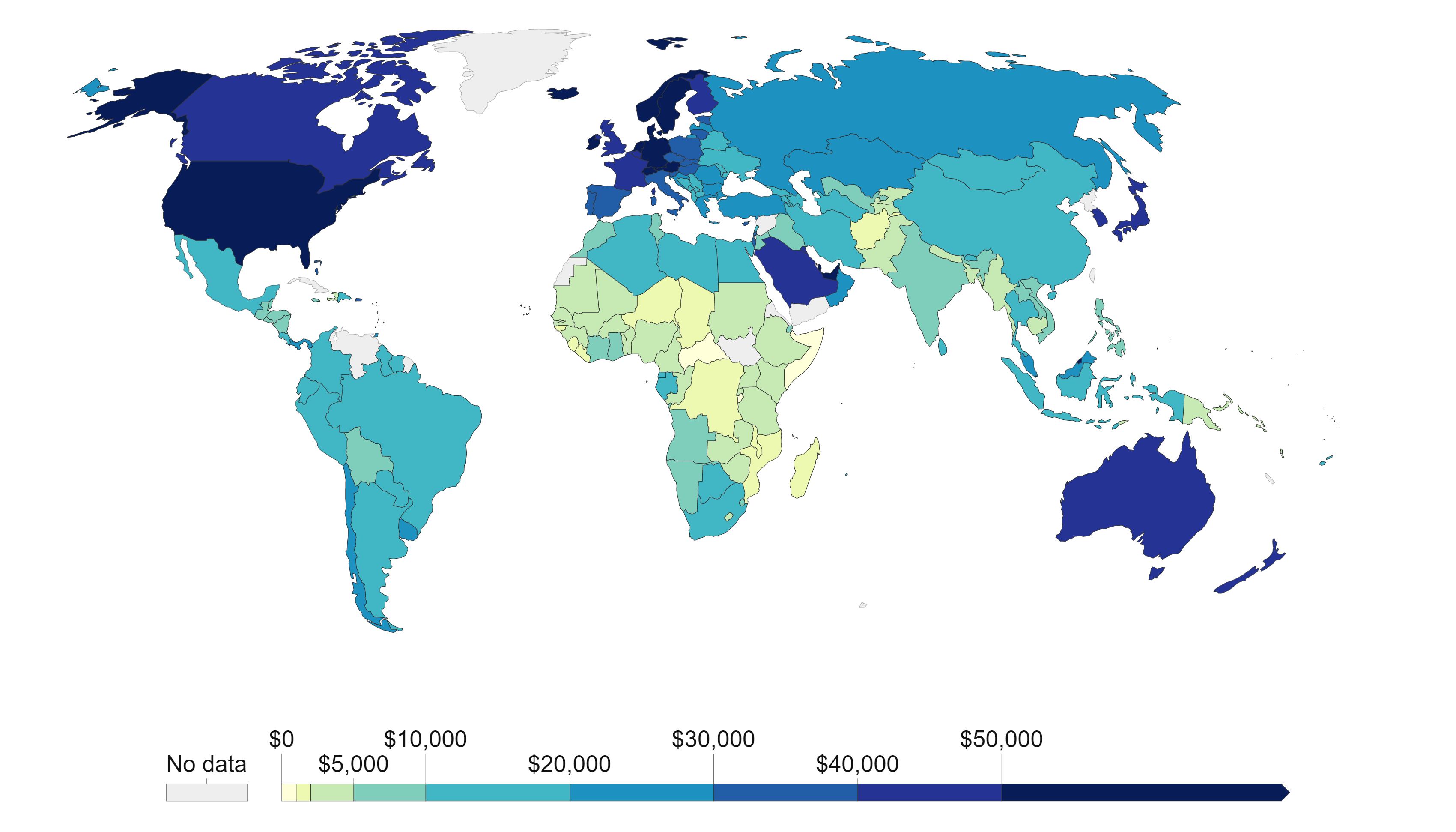}
					\setlength{\abovecaptionskip}{0em}
					\caption{World GDP per Captica}
				\end{figure}
				Let $\gamma$ be the poverty index of the kth country and we specify the value of $\gamma$ could be calculated by equation \ref{equ18} .
				\begin{equation}
					\gamma =
					\left\{
					\begin{aligned}
						&\ 1.2\text{, if the country is one of the 20 countries with the lowest GDP}, \\
						&\ 1.0\text{, otherwise}.
					\end{aligned}
					\right.
					\label{equ18}
				\end{equation}
				
				Then, the formula for total profit is:
				\begin{equation}
					pro_c = \gamma [(income - cost) \cdot \frac{Eq_k}{\sum{Eq_k}}] \\
					\label{equ19}
				\end{equation}
				where $pro_c$ represents the total profit of Country C.
				
\begin{figure}[H]
					\begin{minipage}{0.5\linewidth}
						\centerline{\includegraphics[width=7cm, height=4.2cm]{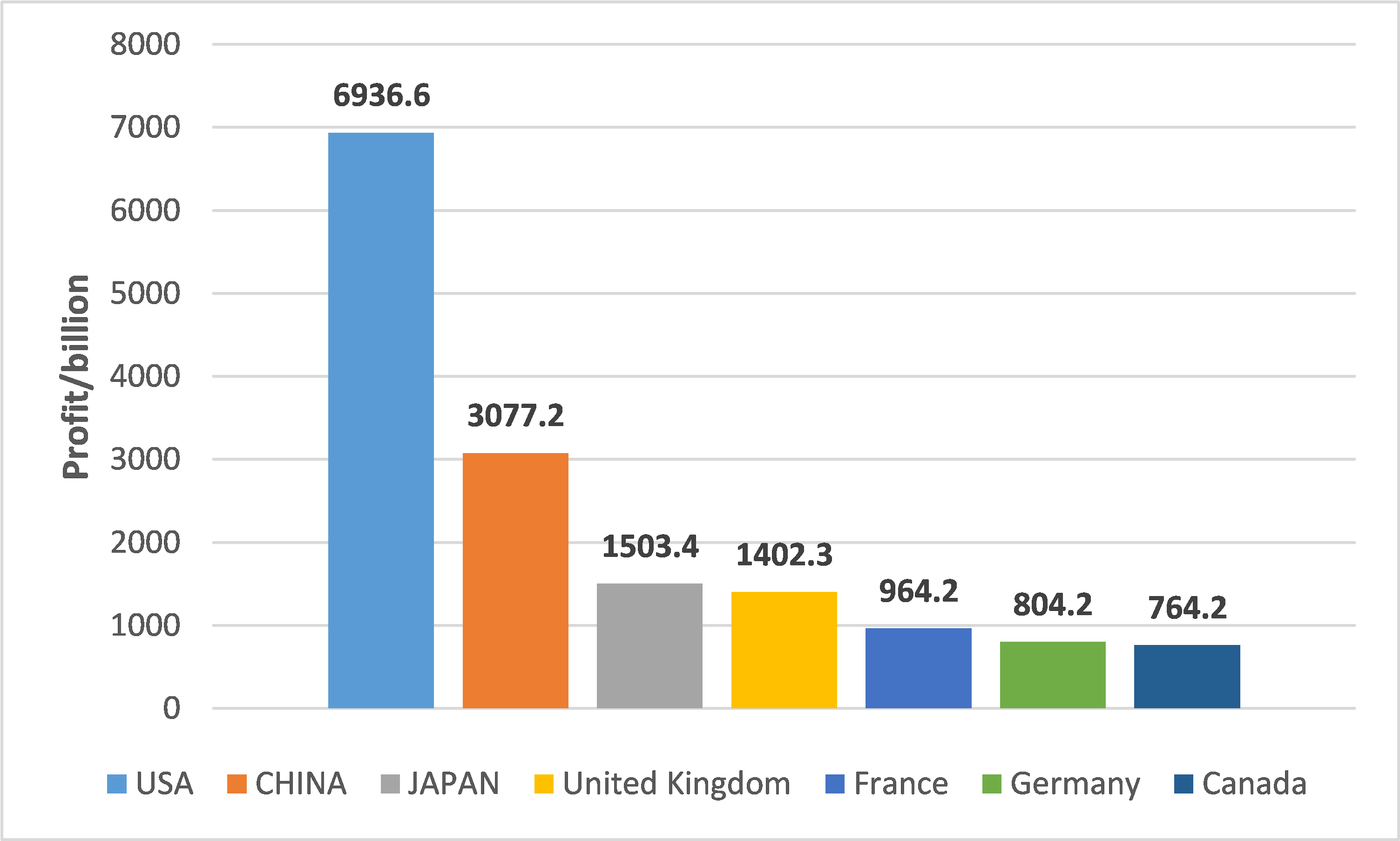}}
						\caption{Top 9 Countries with the Highest Degree of Inequity}
						\label{fig7}
					\end{minipage}
					\hspace{0.2mm}
					\begin{minipage}{0.5\linewidth}
						\centerline{\includegraphics[width=7cm, height=4.2cm]{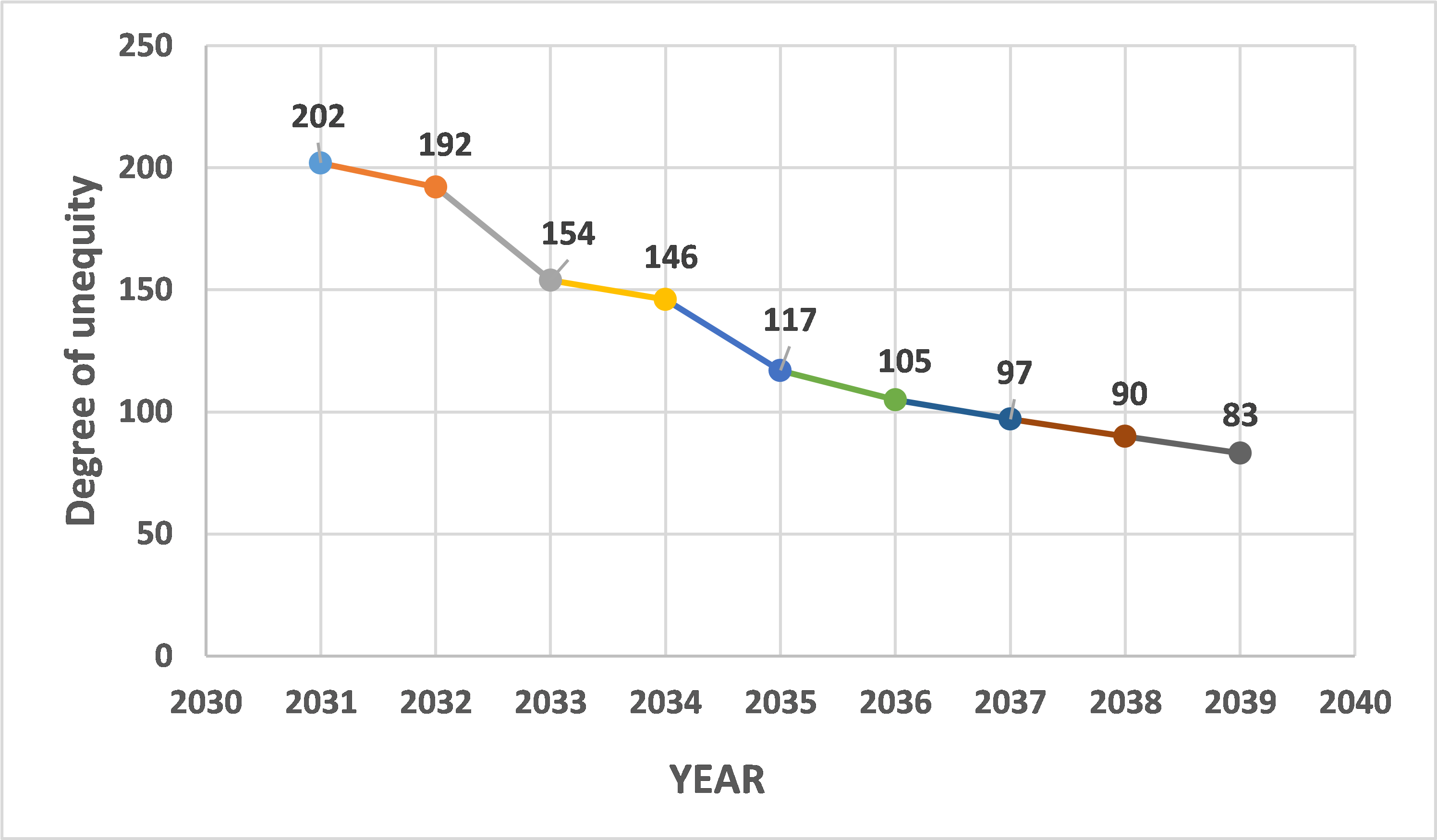}}
						\caption{the Global Inequity Index for 2012 - 2021}
						\label{fig8}
					\end{minipage}
				\end{figure}

				After calculating the total profits for each country in turn using equation \ref{equ11} , we represent the top 7 countries with the highest total profits in figure \ref{fig7} .
				Finally, we calculated the Global Inequity Index for 2030-2039 and the results are shown in figure \ref{fig8} .
				
	\section{Impacts of Changing Conditions on Global Equity}
		To quantitatively measure the impact of each indicator on Country Development Score, we use a back propagation algorithm to calculate it and validate it by Pearson correlation coefficient analysis.
		\subsection{Backward Propagation Algorithm}
			\subsubsection{Brief Introduction to the Back Propagation Algorithm}
				The backpropagation algorithm is a supervised learning method used in conjunction with optimization algorithms such as gradient descent\cite{li2014highly}. It is a generalization of the Delta rule for multilayer feedforward networks, which allows the gradient to be computed using a chain rule for each layer iteration. 
				
				In a word, the following equation holds:
				\begin{equation}
					\left\{
					\begin{aligned}
						z_1^{(l)} =& \ w^{(l)}_{11} a^{(l-1)}_{1} + w^{(l)}_{12} a^{(l-1)}_2 + \cdots \\&+ w^{(l)}_{1N^{(l-1)}} a^{(l-1)}_{N^{(l-1)}} + b^{(l)}_1 \\
						z_2^{(l)} =& \ w^{(l)}_{21} a^{(l-1)}_{1} + w^{(l)}_{22} a^{(l-1)}_2 + \cdots \\&+ w^{(l)}_{2N^{(l-1)}} a^{(l-1)}_{N^{(l-1)}} + b^{(l)}_2 \\
						\vdots &
						\\
						z^{(l)}_{N^{(l)}} =& \ w^{(l)}_{N^{(l)}1} a^{(l-1)}_{1} + w^{(l)}_{N^{(l)}2} a^{(l-1)}_2 + \cdots \\&+ w^{(l)}_{N^{(l)}N^{(l-1)}} a^{(l-1)}_{N^{(l-1)}} + b^{(l)}_{N^{(l)}} \\
					\end{aligned}
					\right. ,
					\label{equ20}
				\end{equation}
				written in the form of matrix multiplication:
				\begin{equation}
					\begin{aligned}
						\begin{bmatrix}
							z^{(l)}_1 \\
							z^{(l)}_2 \\
							\vdots \\
							z^{(l)}_{N^{(l)}} \\
						\end{bmatrix} 
						&=
						\begin{bmatrix}
							w^{(l)}_{11} & w^{(l)}_{12} \cdots & w^{(l)}_{1N^{(l-1)}} \\
							w^{(l)}_{21} & w^{(l)}_{22} \cdots & w^{(l)}_{2N^{(l-1)}} \\
							\vdots \\
							w^{(l)}_{N^{(l)}1} & w^{(l)}_{N^{(l)}2} \cdots & w^{(l)}_{N^{(l)}N^{(l-1)}} \\
						\end{bmatrix}
						\begin{bmatrix}
							a^{(l-1)}_1 \\
							a^{(l-1)}_2 \\
							\vdots \\
							a^{(l-1)}_{N^{(l-1)}} \\
						\end{bmatrix}
						\\&+
						\begin{bmatrix}
							b^{(l)}_1 \\
							b^{(l)}_2 \\
							\vdots \\
							b^{(l)}_{N^{(l)}} \\
						\end{bmatrix},
					\end{aligned}
					\label{equ21}
				\end{equation}
				that is:
				\begin{equation}
					\boldsymbol{z^{(l)}} =  \boldsymbol{w^{(l)}} \boldsymbol{a^{(l-1)}} + \boldsymbol{b^{(l)}}.
					\label{equ22}
				\end{equation}
				Since $\boldsymbol{a^{(l)}} = \sigma (\boldsymbol{z^{(l)}})$, we can conclude that
				\begin{equation}
					\boldsymbol{a^{(l)}} = \sigma  (\boldsymbol{w^{(l)}} \boldsymbol{a^{(l-1)}} + \boldsymbol{b^{(l)}})
					\label{equ23}
				\end{equation}
				where $\sigma(x)$ denotes Activation Function, ususally take $\sigma(x) = \frac{1}{1 + e^{-x}}$.
			\subsubsection{Error Analysis}
\begin{figure}[H]
					\centering
					\includegraphics[width=8.5cm, height=6cm]{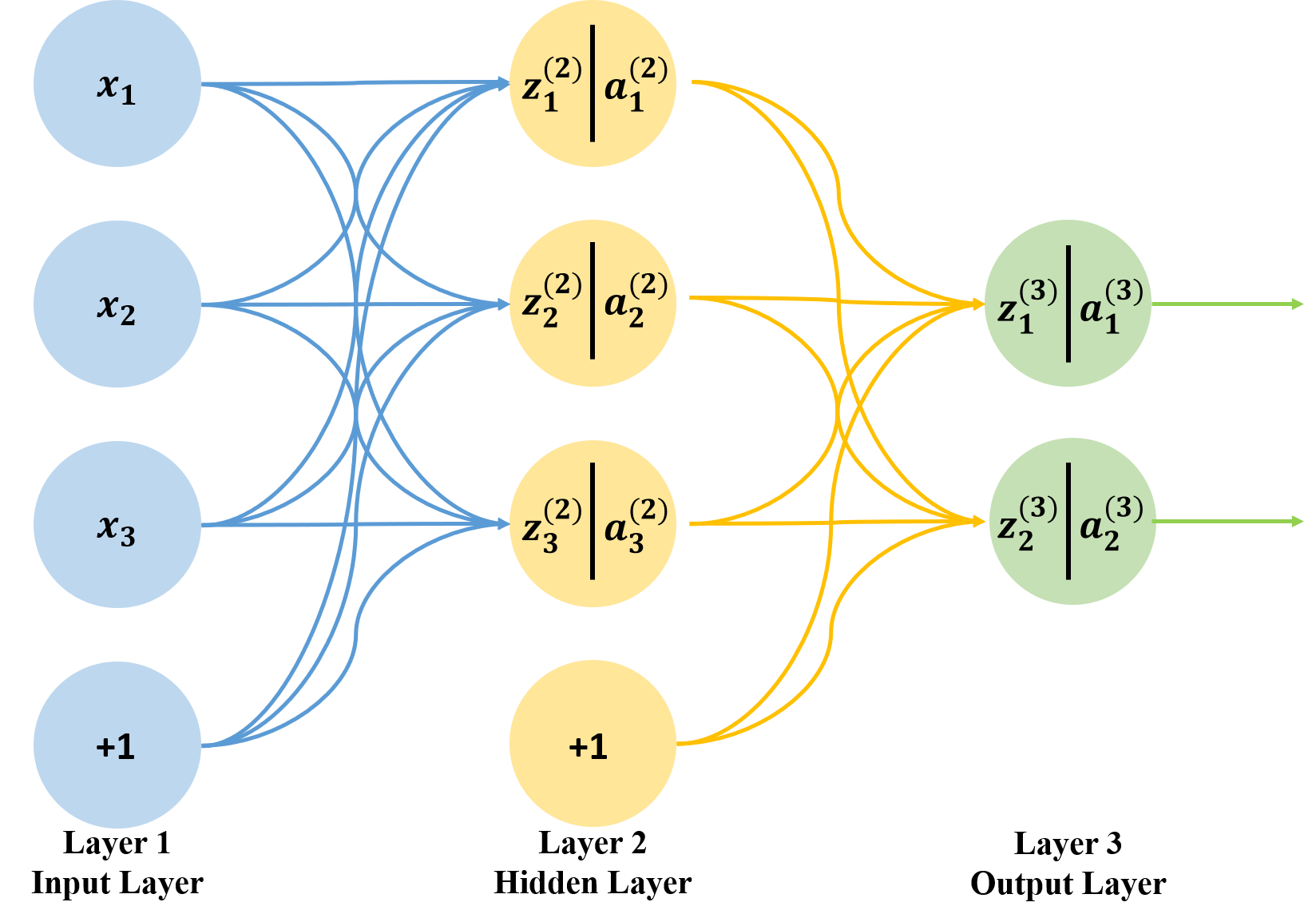}
					\setlength{\abovecaptionskip}{0em}
					\caption{Neural Network Hierarchy}
					\label{fig9}
				\end{figure}

				The back propagation algorithm can be divided into three main layers, namely the input layer, the hidden layer and the output layer, and the relationship between them is shown in figure \ref{fig9} .
				
				{$\divideontimes$ \bfseries{Errors in the output layer}}
				
				According to the transmissibility of error propagation $\Delta z^{(L)}_{j} \rightarrow \Delta a^{(L)}_{j} \rightarrow \Delta C(\theta)$ and combined with the chain derivative rule, we found the error of the loss function on the output layer neurons $\delta^{(L)}_j$.
				\begin{equation}
					\begin{aligned}
						\delta^{(L)}_j &= \frac{\partial C(\theta)}{\partial z^{(L)}_j} = \frac{\partial C(\theta)}{\partial a^{(L)}_j} \frac{\partial a^{(L)}_j}{\partial z^{(L)}_j} \\
						&= \frac{\partial C(\theta)}{\partial a^{(L)}_{j}} \frac{\partial \sigma(z^{(L)}_j)}{\partial z^{(L)}_j} \\
						&= \frac{\partial C(\theta)}{\partial a^{(L)}_j}\sigma^{'}(z^{(L)}_j)
					\end{aligned}
					\label{equ24}
				\end{equation}
				
				For all neurons on the output layer, this can be represented as a vector form.
				\begin{equation}
					\begin{aligned}
						\boldsymbol{\sigma^{(L)}} &=
						\begin{bmatrix}
							\sigma^{(L)}_1 \\
							\sigma^{(L)}_2 \\
							\vdots \\
							\sigma^{(L)}_{N^{(L)}} \\
						\end{bmatrix}
						=
						\begin{bmatrix}
							\frac{\partial  C(\theta)}{\partial a^{(L)}_1} \sigma^{'}(z^{(L)}_1) \\
							\frac{\partial  C(\theta)}{\partial a^{(L)}_2} \sigma^{'}(z^{(L)}_2) \\
							\vdots \\
							\frac{\partial C(\theta)}{\partial  a^{(L)}_{N^L}} \sigma^{'}(z^{(L)}_{N^L}) \\
						\end{bmatrix}
						\\&=
						\begin{bmatrix}
							\frac{\partial  C(\theta)}{\partial a^{(L)}_1} \\
							\frac{\partial  C(\theta)}{\partial a^{(L)}_2} \\
							\vdots \\
							\frac{\partial C(\theta)}{\partial a^{(L)}_{N^L}} \\
						\end{bmatrix}
						\odot
						\begin{bmatrix}
							\sigma^{'}(z^{(L)}_1) \\
							\sigma^{'}(z^{(L)}_2) \\
							\vdots \\
							\sigma^{'}(z^{(L)}_{N^{L}}) \\
						\end{bmatrix} \\
						&= \nabla_{\boldsymbol{a^{(L)}}} C(\theta) \odot \sigma^{'}(\boldsymbol{z^{(L)}}) \\
					\end{aligned}
					\label{equ25}
				\end{equation}
				where $\odot$ is the Hadamard Product(the product of the corresponding elements of the two matrices).
				
				{$\divideontimes$ \bfseries{Error in the Hidden Layer}}
				
				Because the error of the output layer has been found above, according to the principle of error back propagation, the error of the current layer can be understood as a composite function of the error of all neurons in the previous layer(the error of the previous layer is used to represent the error of the current layer, and so on).
				\begin{equation}
					\begin{aligned}
						\sigma^{(l)}_j 
						&= \frac{\partial C(\theta)}{\partial z^{(l)}_j} \\
						&= \sum\limits_{k=1}^{N^{(l+1)}} \frac{\partial C(\theta)}{\partial z^{(l+1)_k}} \frac{\partial z^{(l+1)}_k}{\partial a^{(l)}_j} \frac{\partial a^{(l)}_j}{\partial z^{(l)}_j} \\
						&= \sum\limits_{k=1}^{N^{(l+1)}} \delta^{(l+1)}_{k} \frac{\partial (\sum\limits_{s=1}^{N^{(l)}}w^{(l+1)}_{ks} a^{(l)}_s + b^{(l+1)}_k)}{\partial a^{(l)}_j} \frac{\partial a^{(l)}_j}{\partial z^{(l)}_j} \\
						&= \sum\limits_{k=1}^{N^{(l+1)}} \delta^{(l+1)}_k w^{(l+1)}_{kj} \sigma^{'}(z^{(l)}_j) \\
					\end{aligned}.
					\label{equ26}
				\end{equation}
				
				Similarly for the error of all neurons in the hidden layer, it can be written in vector form.
				\begin{equation}
					\begin{aligned}
						\boldsymbol{\delta^{(l)}} &= 
						\begin{bmatrix}
							\delta^{(l)}_1 \\
							\delta^{(l)}_2 \\
							\vdots \\
							\delta^{(l)}_{N^{(l)}} \\
						\end{bmatrix}
						=
						\begin{bmatrix}
							\sum\limits_{k=1}^{N^{(l+1)}}  \delta^{(l+1)}_{k} w^{(l+1)}_{k1} \sigma^{'}(z^{(l)}_1) \\
							\sum\limits_{k=1}^{N^{(l+1)}} \delta^{(l+1)}_{k} w^{(l+1)}_{k2}  \sigma^{'}(z^{(l)}_2) \\
							\vdots \\
							\sum\limits_{k=1}^{N^{(l+1)}} \delta^{(l+1)}_{k} w^{(l+1)}_{kN^{(l)}}  \sigma^{'}(z^{(l)}_{N^{(l)}}) \\
						\end{bmatrix}
						\\&=
						\begin{bmatrix}
							\sum\limits_{k=1}^{N^{(l+1)}}  \delta^{(l+1)}_{k} w^{(l+1)}_{k1} \\
							\sum\limits_{k=1}^{N^{(l+1)}}  \delta^{(l+1)}_{k} w^{(l+1)}_{k2} \\
							\vdots \\
							\sum\limits_{k=1}^{N^{(l+1)}}  \delta^{(l+1)}_{k} w^{(l+1)}_{kN^{(l)}} \\
						\end{bmatrix}
						\odot
						\begin{bmatrix}
							\sigma^{'}(z^{(l)}_1) \\
							\sigma^{'}(z^{(l)}_2) \\
							\vdots \\
							\sigma^{'}(z^{(l)}_{N^{(l)}}) \\
						\end{bmatrix} \\
						&= (\boldsymbol{w^{(l+1)}})^{T} \boldsymbol{\delta^{(l+1)}} \odot \sigma^{'}(\boldsymbol{z^{(l)}})
					\end{aligned}
					\label{equ27}
				\end{equation}
			
				Through the above steps, we get the results and show them in table \ref{tab12} .
				\begin{table}[H]
					\centering
					\arrayrulecolor{ruleblue}
					\setlength{\abovecaptionskip}{0em}
					\caption{Results of Backward Propagation Algorithm}
					\label{tab12}
					\begin{tabular}{cp{1.2cm}<{\centering}p{1.2cm}<{\centering}p{1.2cm}<{\centering}p{1.4cm}<{\centering}p{1.2cm}<{\centering}p{1.2cm}<{\centering}p{1.2cm}<{\centering}}
						\Hline
						\rowcolor{rowheadblue}
						\textbf{\textit{\color{white}Indicators}} 
						& \textbf{\textit{\color{white}EI}} & \textbf{\textit{\color{white}IDG}}& \textbf{\textit{\color{white}CEA}}& \textbf{\textit{\color{white}MA}} & \textbf{\textit{\color{white}HR}} & \textbf{\textit{\color{white}ER}} & \textbf{\textit{\color{white}SA}} \\
						\Hline
						Value & 0.042 & 0.071 & 0.029 & 0.013 & 0.015 & 0.021 & 0.019 \\
						\Hline
					\end{tabular}
				\end{table}
		\subsection{Pearson Correlation Analysis}
			\subsubsection{Basic Principles}
				Pearson correlation analysis is used to explore the correlation between world equity scores and each representative indicator of each country. The Pearson correlation coefficient between the two variables is defined by the equation \ref{equ28}:
				\begin{equation}
					\rho_{(Eq_k, GDP)} = \frac{cov(Eq_k, GDP)}{\sigma_{Eq_k} \sigma_{GDP}} = \frac{E[(Eq_k - \overline{Eq_k})(GDP - \overline{GDP})]}{\sigma_{Eq_k} \sigma_{GDP}}
					\label{equ28}
				\end{equation}
				
				Then we use equation \ref{equ29} to calculate the correlation coefficient r.
				\begin{equation}
					r = \frac{\sum\limits_{i = 1}^{n}(Eq_k - \overline{Eq_k})(GDP - \overline{GDP})}{\sqrt{\sum\limits_{i = 1}^{n}(Eq_k - \overline{Eq_k})^2} \sqrt{\sum\limits_{i = 1}^{n}(GDP - \overline{GDP})^2}}
					\label{equ29}
				\end{equation}
				
\begin{figure}[H]
					\centering
					\includegraphics[width=\textwidth, height=3.5cm]{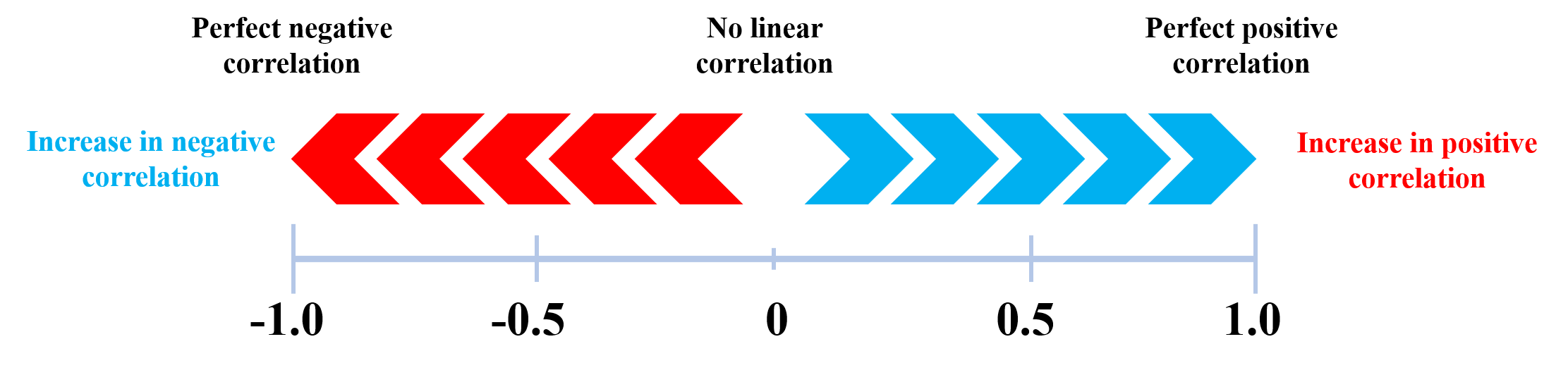}
					\setlength{\abovecaptionskip}{0em}
					\caption{Judgment of the Degree of Relevance}
					\label{fig10}
				\end{figure}

				Next, the correlation between the two variables is determined based on the r values. And the priciples are shown in figure \ref{fig10} .
				As can be seen from figure \ref{fig10} , the closer the absolute value of r is to 1, the stronger the correlation between the two variables. Meanwhile, the positive or negative of r determines the positive or negative of the correlation.
				
			\subsubsection{Calculation Process}
				{$\divideontimes$ \bfseries{Calculation of Correlation Coefficient}}
				
				Take analysis of the correlation between GDP and countries' equity scores as an example, through the analysis of the first question we know that the mean value of $Eq_k$ is 34.7227. Then, we use equation \ref{equ30} .
				\begin{equation}
					\begin{aligned}
						\sigma_{Eq_k} &= \sqrt{\sum\limits (Eq_k  - \overline{Eq_k})^2} \\
						&= \sqrt{\sum\limits (Eq_k - 34.7227)^2} \\
						&= 5.7147
					\end{aligned}
					\label{equ30}
				\end{equation}
				to got the variance is 32.6574 and the standard deviation is 5.7147.
				
				Similarly, since the average value of GDP is 11540.8620, we substitute it into equation \ref{equ31} .
				\begin{equation}
					\begin{aligned}
						\sigma_{GDP} &= \sqrt{\sum (GDP - \overline{GDP})^2} \\
						&= \sqrt{\sum (GDP - 11540.8620)^2} \\
						&= 11.9809
					\end{aligned}
					\label{equ31}
				\end{equation}
				that is, ehile variance is 143.5418, the standard deviation is 11.9809.
				
				Finally, we calculate the sum of the outlying product of the score and GDP by equation \ref{equ32} .
				\begin{equation}
					\begin{aligned}
						cov(Eq_k, GDP) &= \sqrt{\sum\limits (Eq_k  - \overline{Eq_k})(GDP - \overline{GDP})} \\
						&= \sqrt{\sum\limits (Eq_k  - 34.7227)(GDP - 11540.8620)} \\
						&= 6.6509.
					\end{aligned}
					\label{equ32}
				\end{equation}
				
				The sum of the outlying product of the score and GDP is 6.6509.
				Substitute into equation \ref{equ33}.
				\begin{equation}
					\begin{aligned}
						\rho_{(Eq_k, GDP)} &= \frac{cov(Eq_k,  GDP)}{\sigma_{Eq_k} \sigma_{GDP}} = \frac{\sum\limits_{i = 1}^{n}\frac{(Eq_k - \overline{Eq_k})}{\sigma_{Eq_k}}\frac{(GDP - \overline{GDP})}{\sigma_{GDP}}}{n} \\
						&= \frac{\sum\limits_{i = 1}^{n}\frac{(Eq_k - 34.7227)}{\sigma_{Eq_k}}\frac{(GDP - 11540.8620)}{\sigma_{GDP}}}{n}\\
						&= 0.871
					\end{aligned}
					\label{equ33}
				\end{equation}
				
				{$\divideontimes$ \bfseries{Test of Pearson's Correlation Coefficient}}
				
				After obtaining the correlation coefficient, we use the Pearson Correlation Coefficient Method to test it which using equation \ref{equ34} .
				\begin{equation}
					\begin{aligned}
						r &= \frac{\sum\limits_{i = 1}^{n}(Eq_k - 34.7227)(GDP - 11540.8620)}{\sqrt{\sum\limits_{i = 1}^{n}(Eq_k - 34.7227)^2} \sqrt{\sum\limits_{i = 1}^{n}(GDP - 11540.8620)^2}} \\
						&= 0.78.
					\end{aligned}
					\label{equ34}
				\end{equation}
				
				Next, we propose the hypothesis:
				
				{$\bullet$ $H_0: P = 0$, score is not related to GDP;}
				
				{$\bullet$ $H_1: P \neq 0$, score is not related to GDP;}
				
				Meanwhile, we determine the corresponding significant level of 0.05.
				
				Using equation \ref{equ35} , we obtain the value of r.
				\begin{equation}
					\begin{aligned}
						t_r &= \frac{|r - 0|}{\sqrt{(1 - r^2)/(n - 2)}} \\
						&= \frac{0.78}{\sqrt{(1-0.78^2)/(7-2)}} \\
						&= 19.5894
					\end{aligned}
					\label{equ35}
				\end{equation}
				
				Since we check the t-test adjacency table, we have obtain the threshold $P = 0.9$ and the linear correlation coefficient is 1.653. As a result, we accept the original hypothesis and deem that GDP has a significant impact on the global equity.
				\begin{table}[H]
					\centering
					\arrayrulecolor{ruleblue}
					\renewcommand\arraystretch{1}
					\setlength{\abovecaptionskip}{0em}
					\caption{T-test Adjacency Table}
					\begin{tabular}{p{2.5cm}<{\centering}p{3cm}<{\centering}p{3cm}<{\centering}p{3cm}<{\centering}}
						\Hline
						\rowcolor{rowheadblue}
						\textbf{\textit{\color{white}\diagbox{n}{P}}} & \textbf{\textit{\color{white}0.25}} & \textbf{\textit{\color{white}0.1}} & \textbf{\textit{\color{white}0.05}} \\
						\Hline
						100 & 0.677 & 1.290 & 1.660 \\
						\Hline
						\rowcolor{rowblue}
						200 & 0.676 & 1.653 & 1.972 \\
						\Hline
						500 & 0.675 & 1.283 & 1.648 \\
						\Hline
					\end{tabular}
				\end{table}
	\section{Sensitivity Analysis}
\begin{figure}[H]
			\centering
			\includegraphics[width=8.5cm, height=6cm]{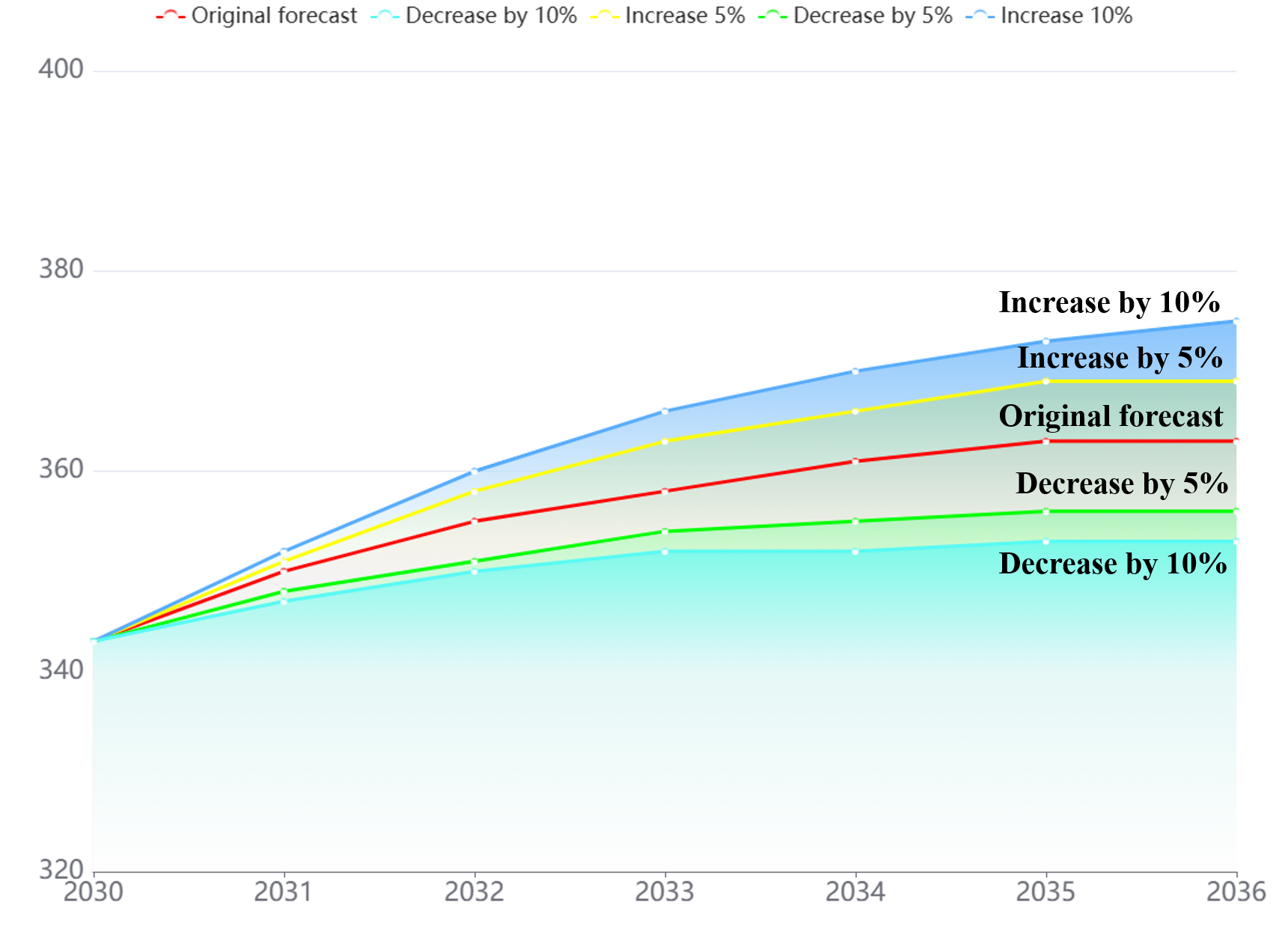}
			\setlength{\abovecaptionskip}{0em}
			\caption{Sensitivity Analysis}
			\label{fig11}
		\end{figure}

		According to the transmissibility of error propagation, $\Delta w^{(l)}_{jk} \rightarrow \Delta z^{(j)}_{j} \rightarrow \cdots \rightarrow \Delta C(\theta)$, the loss function can be viewed as a composite function of the weights ($w$). From the chain rule of derivation
		\begin{equation}
			\begin{aligned}
				\frac{\partial C(\theta)}{\partial w^{(l)}_{jk}} &= \frac{\partial C(\theta)}{\partial z^{(l)}_{j}} \frac{\partial z^{(l)}_{j}}{\partial w^{(l)}_{jk}} \\
				&= \delta^{(l)}_{j} \frac{\partial (\sum\limits_{s=1}^{N^{(l-1)}}w^{(l)}_{js} a^{(l-1)}_{s} + b^{(l)}_{s})}{\partial w^{(l)}_{jk}} \\
				&= \delta^{(l)}_{j} a^{(l - 1)}_{k} \\
				\Delta Z &= \int_{w_1}^{w_2} \delta^{(l)}_{j} a^{(l - 1)}_{k} dw,
			\end{aligned}
			\label{equ36}
		\end{equation}
		we have calculated the results and shown them in figure \ref{fig11} .
		The results show that our model is stable, with a variation of no more than 7\%.
		
		\bibliographystyle{ieeetr}
		\bibliography{article}

\end{document}